\documentclass[twocolumn,showpacs,preprintnumbers,nofootinbib,prd,superscriptaddress,10pt]{revtex4-1}
\usepackage{graphicx,amssymb,amsmath,amsthm,amsfonts,epsfig}

\usepackage[linktocpage]{hyperref}
\usepackage[usenames,dvipsnames]{color}
\usepackage{epstopdf}
\usepackage{aas_macros}
\usepackage{pifont}
\definecolor{darkred}{rgb}{0.5,0,0}
\definecolor{darkgreen}{rgb}{0,0.5,0}
\definecolor{darkblue}{rgb}{0,0,0.5}
\definecolor{prussian}{rgb}{0.0, 0.19, 0.33}
\definecolor{richelectricblue}{rgb}{0.03, 0.57, 0.82}
\definecolor{teal}{rgb}{0.0, 0.5, 0.5}
\definecolor{mediumseagreen}{rgb}{0.24, 0.7, 0.44}
\definecolor{lust}{rgb}{0.9, 0.13, 0.13}
\definecolor{ballblue}{rgb}{0.13, 0.67, 0.8}
\definecolor{darkcyan}{rgb}{0.0, 0.55, 0.55}
\definecolor{mountainmeadow}{rgb}{0.19, 0.73, 0.56}
\definecolor{palecarmine}{rgb}{0.69, 0.25, 0.21}
\definecolor{richcarmine}{rgb}{0.84, 0.0, 0.25}
\definecolor{tangelo}{rgb}{0.98, 0.3, 0.0}
\definecolor{venetian}{rgb}{0.784,0.031,0.082}
\definecolor{bdfrance}{rgb}{0.192,0.549,0.906}

\hypersetup{colorlinks=true, citecolor=venetian,
linkcolor=bdfrance, urlcolor=blue}
\usepackage{amsmath,amssymb}
\usepackage{tensor}
\usepackage{mathtools}
\usepackage{amsbsy}
\usepackage{bm}
\usepackage{float}
\newcommand{\be}{\begin{equation}}
\newcommand{\ee}{\end{equation}}
\newcommand{\bear}{\begin{eqnarray}}
\newcommand{\eear}{\end{eqnarray}}

\newcommand{\p}{\prime}
\newcommand{\pp}{\prime\prime}

\newcommand{\cZ}{{\cal Z}}

\newcommand{\cS}{{\cal S}}
\newcommand{\cR}{{\cal R}}
\newcommand{\cF}{{\cal F}}

\newcommand{\cB}{{\cal B}}
\newcommand{\cA}{{\cal A}}
\newcommand{\cX}{{\cal X}}

\newcommand{\rZ}{{\rm Z}}

\newcommand{\rT}{{\rm T}}
\newcommand{\rC}{{\rm C}}
\newcommand{\rD}{{\rm D}}
\newcommand{\rRW}{{\rm RW}}

\newcommand{\rp}{r_{+}}
\newcommand{\tx}{\tilde{x}}

\begin{document}

%\preprint{APS/123-QED}

\title{The Darboux transformation in black hole perturbation theory}% Force line breaks with \\

\author{Kostas Glampedakis}
\email{kostas@um.es}
\affiliation{Departamento de F\'isica, Universidad de Murcia,
Murcia, E-30100, Spain}
\affiliation{Theoretical Astrophysics, University of T\"ubingen, Auf der Morgenstelle 10, T\"ubingen, D-72076, Germany}

\author{Aaron D. Johnson}
 \affiliation{%
Department of Physics, University of Arkansas, Fayetteville, Arkansas 72701, USA
 }%

\author{Daniel Kennefick}
\affiliation{%
Department of Physics, University of Arkansas, Fayetteville, Arkansas 72701, USA
}%
\affiliation{%
Arkansas Center for Space and Planetary Sciences, University of Arkansas, Fayetteville, Arkansas 72701, USA
}%

\noaffiliation

\date{\today}
% It is always \today, today,
%  but any date may be explicitly specified

\begin{abstract}

The Darboux transformation between ordinary differential equations is a 19th century technique
that has seen wide use in quantum theory for producing exactly solvable potentials for the Schr\"odinger
equation with specific spectral properties. In this paper we show that the same transformation appears in
black hole theory, relating, for instance, the Zerilli and Regge-Wheeler equations for 
axial and polar Schwarzschild perturbations. The transformation reveals these two equations to be
isospectral, a well known result whose method has been repeatedly reintroduced 
under different names. We highlight the key role that the so-called algebraically special solutions 
play in the black hole Darboux theory and show that a similar relation exists between the 
Chandrasekhar-Detweiler equations for Kerr perturbations. Finally, we discuss the limitations of the method
when dealing with long-range potentials and explore the possibilities offered by a generalised Darboux
transformation. 

\end{abstract}

% PACS, the Physics and Astronomy Classification Scheme.
%\keywords{Suggested keywords}
%Use showkeys class option if keyword display desired
\maketitle

%\tableofcontents

%%%%%%%%%%%%%%%%%%%%%%

\section{Introduction}
\label{sec:intro}

In the 1970s Chandrasekhar found a curious relationship between the equations 
governing black hole perturbations of different parity. The Regge-Wheeler 
equation\,\cite{RW57}, which governs odd-parity, or axial, perturbations and the Zerilli 
equation\,\cite{zerilli70}, which governs even-parity, or polar perturbations, are isospectral. 
There exists a transformation, known as the Chandrasekhar transformation, 
which relates the solutions of the two equations to each other\,\cite{chandrabook}. Each solution 
of the one can be transformed into one of the solutions of the other, so that 
each solution of the Regge-Wheeler equation corresponds to a solution of the 
Zerilli equation. This characteristic of isospectrality can be very useful, 
since there may be occasions when it would be easier to solve an equation other 
than the one which more correctly describes the situation one is interested in. 
Chandrasekhar himself regarded it as a great mystery that these equations 
should be isospectral. When one of us visited him in 1995 he gave it as an 
example of an outstanding issue in the field which was, he felt, treated as 
merely a useful curiosity when it should be investigated more deeply. 

Of course the matter has been investigated many times, by Chandrasekhar and
others, but still more with a view to exploiting its utility than to elucidating the mystery. 
For instance, it has been discovered that a similar situation exists in the case of Kerr 
black holes. The Teukolsky equation which describes perturbations of rotating 
black holes has a long range potential\,\cite{teuk73}. 
In 1982 Sasaki and Nakamura discovered a new equation, named after them, 
which is isospectral with the Teukolsky equation but has a shorter ranged 
potential\,\cite{SN82}. They gave a transformation which permits one to go from solutions 
of the Sasaki-Nakamura equation to equivalent solutions of the Teukolsky 
equation\,\cite{SN82}.

In the spirit of Chandrasekhar's advice to delve into the reasons behind this 
interesting isospectral property of black-hole perturbations, we draw 
attention to the fact that the Chandrasekhar transformation is an example of 
the Darboux transformation. 

Discovered in the late 19th century by the French 
mathematician Jean-Gaston Darboux\,\cite{darboux}, this transformation is well known to applied 
mathematicians and quantum physics practitioners for its usefulness, amongst a range of applications, 
in relating solutions of different potentials in scattering problems and generating new exactly solvable potentials 
for the Schr\"odinger equation from known ones. An excellent introduction to the subject can be found in\,\cite{greekbook}.
Chandrasekhar himself noted the similarity of his transformation with techniques used in studying the 
Korteweg-de Vries (KdV) equation\,\cite{chandrabook} and it is not surprising that the Darboux 
transformation plays an important role in this field. 
 It is also central to the theory of supersymmetry where it is 
the primary technique used to find supersymmetric partner solutions to the existing solutions of the standard 
model\,\cite{purseyII, review2}.

In this paper we explore the Darboux transform as it applies to gravitational 
waves scattering off black holes. In particular, we investigate methods by 
which one might discover new isospectral equations which would be useful in 
future studies. We also examine methods which are useful in tackling the sometimes daunting 
complexity of the Darboux method. For instance algebraically special solutions of the perturbation 
equations, in which there is no backscatter from a wave impinging on a black 
hole, provide an easier route to discovering the Darboux transformation between
one potential and another. We also show that the Darboux transformation cannot 
be used to transform a long-ranged potential into a short-ranged potential. 
Thus the Sasaki-Nakamura transform is not of the Darboux type. Instead it is an 
example of a generalized Darboux transformation,\cite{darboux}. In this paper we will focus on 
which of the many different equations describing black hole potentials are 
related by the standard Darboux transform. Not all of them are. Some are
related instead by the generalized version. Nor is the Darboux method the only one known to be useful in 
looking for isospectral potentials. Other methods are used, for instance, in looking for solutions 
of the Schr\"odinger equation\,\cite{abmoses}. Nevertheless we feel that it may be useful, in black
hole perturbation theory, to make note of the role played by transformations
of the Darboux type, in the hope that a better understanding of the 
mechanisms which can be used to generate these kinds of relationships will
make finding useful equations of the Sasaki-Nakamura type easier in the
future.

Of course there have been many previous attempts to understand the mystery of
Chandrasekhar's isospectral transformations. 
The earliest example is the paper by Heading\,\cite{heading} which has shown how the different Schwarzschild black hole 
perturbation equations are linked by a generalized type of transformation. 
Meanwhile, authors such as Anderson and Price\,\cite{ap91}  have attempted to introduce a terminology 
(referring to the use of  ``inter-twining operators") and van den Brink has commented upon the 
similarity between such techniques and those used in supersymmetry theory\,\cite{maassen00}. 
These earlier treatments, however, do not draw the connection to the Darboux family of transformations. 
We feel it is important to introduce this terminology in order to facilitate drawing
useful lessons from the use of the Darboux method in other fields of physics
and mathematics. 
A source of useful references in this broader literature is an article by Luban and Pursey\,\cite{LP86} 
which shows that the Darboux transformation is not the only method for finding isospectral potentials, citing the method
of Abraham and Moses (based on the Gel'fand-Levitan equation)\,\cite{abmoses}.
Luban and Pursey also give a useful survey of different applications of the Darboux transformation, 
a notable example dating back to Schr\"odinger himself and related to the celebrated factorisation method
in quantum mechanics\,\cite{schroedinger41,IH51}.
One quickly learns from their paper that using the Darboux method while remaining unaware of Darboux's 
papers is not confined to the field of black hole perturbation theory.

The reminder of this paper is organised as follows. 
In Section~\ref{sec:darboux} we provide an introduction to the Darboux transformation and describe 
some equations which are useful in identifying potentials which are Darboux-related. 
In Section~\ref{sec:darbouxS} we discuss several aspects
of the Darboux theory as it applies to Schwarzschild black holes, including
the original Chandrasekhar transformation (Subsections~\ref{sec:scptb} and ~\ref{sec:darbouxRW}), the 
importance of the algebraically special reflectionless solution (Subsection~\ref{sec:2special}) and 
the interesting fact that the long-ranged nature of potential in the
Bardeen-Press equation means that it cannot be Darboux related to the
Regge-Wheeler and Zerilli potentials, which are short-ranged (Subsections~\ref{sec:BP} 
and ~\ref{sec:gdarboux}). Finally, in Section~\ref{sec:darbouxK} we discuss the Darboux theory of Kerr
black holes. We examine a group of different Kerr wave equations discovered
by Chandrasekhar and Detweiler and show which of these are Darboux-related
(Subsection~\ref{sec:CDtetrad}) and discuss the use of the algebraically special solutions
in finding new Darboux-transformed wave equations for the Kerr metric
(Subsection~\ref{sec:Kspecial}). We conclude (Section~\ref{sec:conclusions}) with a summary of our results 
and some remarks on possible further applications to the study of black holes and relativistic stars.

Throughout the paper we assume a $e^{-i\omega t}$ time-dependence for the 
perturbations and use geometric units $G=c=1$.

%%%%%%%%%%%%%%%%%%%%%%%%%%%%

\section{The Darboux Transformation}
\label{sec:darboux}

The Darboux transformation (hereafter DT)\,\cite{darboux} is an elegant method for relating second-order 
ordinary differential equations written in `canonical' form, i.e. without first-order derivative terms.

The starting point of the method is an equation of the form (a prime denotes differentiation with respect to 
the argument)
\be
y^{\pp} (x) + q(x) \, y(x) = 0,
\label{yeq0}
\ee
where $q(x)$ is the `initial' potential. We seek to find another equation
in the canonical form
\be
Y^{\pp} (x) + Q(x) \, Y(x) = 0,
\label{Yeq0}
\ee
where $Q(x) $ is the `final' potential. 

The DT between the two equations consists of the linear relation
which relates solutions of the two equations 
\be
Y = y^\p + f(x)y.
\label{darb0}
\ee
Differentiating this twice and  using equation (\ref{yeq0}) and (\ref{darb0}) to eliminate the derivatives 
of $y$ leads to
\be
Y'' = (q - 2f') \, Y + (f^2 + q - f')' \, y = 0.
\ee
Therefore, the DT yields two constraint equations,
\be
Q = q - 2f^\p,
\label{potentials}
\ee
relating the initial and final potentials, and a Riccati equation for the $f$-function:
\be
f' - f^2 - q = c \qquad ( c = \mbox{const.} )
\label{riccati}
\ee
This can be transformed into a linear equation using the standard logarithmic derivative trick:
\be
f = - \frac{u^\p}{u} \quad \Rightarrow \quad u^{\pp} + (\, q + c \, ) u = 0.
\label{ueq0}
\ee
Thus any particular solution of this $u$-equation can serve as a DT 
generator.

To what extent the new potential $Q$ retains a qualitative resemblance  to the initial potential $q$ 
largely depends on the chosen $u$ solution. For instance, a $u(x)$ that vanishes at finite values of $x$
will generate a singular $Q$ (see (\ref{potentials})) even if $q$ is perfectly regular. Hence, if for example
one wishes to go from a barrier-like potential $q$ to a new barrier-like potential $Q$ the chosen $u(x)$ 
should be non-vanishing.

In the particular case where (\ref{yeq0}) \& (\ref{Yeq0}) are Schr\"odinger-type equations, 
\be
y^{\pp} + [\, \lambda - v(x)\,] y = 0, \quad
Y^{\pp} + [\, \lambda - V(x)\,] y = 0,
\ee
with  eigenvalue $\lambda $, the DT becomes
\be
Y = y^\p - \left ( \frac{u^\p}{u} \right ) y, \qquad V = v - 2 \left ( \frac{u^\p}{u} \right )^\p,
\ee
and
\be
u^{\pp} + [\, \mu - v(x) \, ] u = 0.
\ee
As evident from this last expression, the auxiliary function $u$ is a solution of the original equation, albeit with a 
different eigenvalue $\mu = \lambda + c$.

On some occasions one may know the two potentials $q, Q$ and desire to check if they are Darboux-related. 
In that case the $u$-equation (\ref{ueq0}) is of no real use and one can simply check the veracity of the two 
constraint equations (\ref{potentials}) \& (\ref{riccati}). For that purpose these equations can be combined and 
put into a more practical form:
\begin{align}
f = \frac{(Q+q)'}{2(Q-q)},
\label{feq}
\\
\nonumber \\
f^\p = -\frac{1}{2}(Q - q).
\label{dfeq}
\end{align}
These equations completely specify the transformation. It is easy to see that if a given $f$-function 
generates the  $q \rightarrow Q$ DT, the inverse transformation $Q \rightarrow q$  is generated  
by $-f$.

Furthermore, $f$ can be eliminated between the two basic equations (\ref{feq}) and (\ref{dfeq})
and one can arrive at a constraint equation between the two potentials:
\be
\frac{(Q+q)'}{Q-q} = \int dx (Q-q).
\label{Qqeq}
\ee
As we shall see below, this form can be more advantageous than the original constraints
when dealing with potentials that are rational functions.

The DT is a versatile technique for manipulating equations of the form (\ref{yeq0}).
Given that the Schr\"odinger equation falls into that category it is not surprising that the method 
has been widely used in quantum physics see e.g.\,\cite{greekbook,review1,review2}. 
As  already pointed out in the introduction, the transformation allows one to 
produce new exactly solvable potentials from an initial solvable potential and
make specific modifications in a system's energy levels. An equally exciting possibility 
is the Darboux-pairing of potentials that share identical spectral properties
(i.e. transmission/reflection coefficients and energy levels).

The form (\ref{yeq0}) is also routinely produced by wave equations after separation of variables. 
For example, this is the case for the equations governing the dynamics of linear perturbations 
in black hole spacetimes (the most important of these are summarised in Appendix~\ref{sec:canonical}).  
As we discuss in the following sections, some of these wave equations are Darboux-related -- a property 
first noticed by Chandrasekhar in the 1970s but never fully recognised as a DT and which, inevitably, 
led to the reinvention of the method by others\,\cite{heading, ap91, leung99, maassen00}. 
The DT connection then provides a natural explanation for the identical quasinormal mode (QNM) spectra 
(a QNM frequency corresponds to a wave solution that  is purely outgoing (ingoing) at spatial infinity 
(event horizon), see \cite{KKreview, EBreview} for reviews on the subject) and scattering amplitudes between 
certain black hole potentials (the isospectral property of the DT is discussed further in Appendix~\ref{sec:isospectral}).

%%%%%%%%%%%%%%%%%%%%%%%%%%%%%%%%%%%%

\section{Darboux theory for Schwarzschild black holes}
\label{sec:darbouxS}

\subsection{The `mysterious' relation between axial and polar perturbations}
\label{sec:scptb}

The mathematics of black holes is a wonderland  of ``miraculous"  properties such as the 
separability of the linear perturbation equations (leading to the celebrated Teukolsky equation\,\cite{teuk73}) 
and the integrability of geodesic motion (leading to the Carter constant third integral of motion\,\cite{carter68}).  

In his pioneering work on black hole perturbation theory Chandrasekhar stumbled upon one more remarkable 
property\,\cite{chandrabook}:  despite being governed by seemingly different wave equations, the axial and 
polar perturbations of a Schwarzschild black hole have \emph{identical} reflection and transmission amplitudes  
and QNM spectra.

The polar (even-parity) metric perturbations are fully specified by the solution of the 
Zerilli equation\,\cite{zerilli70}
\be
\frac{d^2 Z}{dx^2} + (\, \omega^2 - V_\rZ  \, ) Z = 0,
\label{zerilli}
\ee
where we have used $x$ (instead of the usual $r_*$) to denote the standard tortoise coordinate  
and $\omega$ is the frequency. The Zerilli potential is given by
\begin{multline}
V_\rZ (r) = \frac{(1-2M/r)}{r^3 ( nr + 3M)^2} \left [\, 2 (n+1) n^2 r^3 + 6 n^2 M r^2 \right.
\\
\left. + 18 n M^2 r + 18 M^3  \, \right ],
\label{Vz}
\end{multline}
and the perturbation's multipole order $\ell$ appears in the combination
\be
n = \frac{1}{2} (\ell-1) (\ell+2). 
\ee

The axial (odd-parity) perturbations are described by the much simpler Regge-Wheeler 
(R-W) equation\,\cite{RW57}:
\be
\frac{d^2 X}{dx^2} + (\, \omega^2 - V_\rRW  \, ) X = 0,
\label{rw}
\ee
where
\be
V_\rRW (r) = \left (1-\frac{2M}{r} \right ) \left [\, \frac{\ell (\ell+1)}{r^2} - \frac{6M}{r^3} \, \right ].
\label{Vrw}
\ee

In a \emph{tour de force} calculation Chandrasekhar\,\cite{chandra75} was able to discover the following 
differential relation between the Zerilli and R-W wave functions:
\be
X = A Z^\p + B Z,
\label{trans1}
\ee
where
\begin{align}
A &= -M \left [ i\omega M + \frac{1}{3} n(n+1) \right ]^{-1},
\label{Acoeff}
\\
\nonumber \\
B(r) &= \frac{ n(n+1) (nr+3M) r^2 + 9M^2 (r-2M)}{r^2 (nr+3M) [n(n+1) + 3 i\omega M]}.
\label{Bcoeff}
\end{align}
This relation was subsequently used by Chandrasekhar \& Detweiler to demonstrate 
the equivalence between the axial and polar QNM spectra\,\cite{cd75} and eventually
culminated in the formulation of a black hole transformation theory\,\cite{chandra80},\cite{chandrabook}. 
As already discussed in the introduction, later work demonstrated that Chandrasekhar's formalism 
is a special case of a more general framework\,\cite{heading,ap91,leung99, maassen00} but, somewhat
surprisingly, failed to identify the appearance of the 19th century DT in black hole perturbation theory.

%%%%%%%%%%%%%%%%%%%%%%%%%%%%%%%%%%%%%%%%%%%%%%

\subsection{The Darboux transformation between the Regge-Wheeler and Zerilli equations}
\label{sec:darbouxRW}

In this section we show that the Chandrasekhar transformation (\ref{trans1}) between the R-W and 
Zerilli equations is nothing else than an ordinary DT. 
In particular, it is shown to be a DT generated by a $u$ function [see Eq.~(\ref{ueq0})] that corresponds 
to an \emph{algebraically special} solution of the R-W equation (the key role of this solution
in this context has already been pointed out in Refs.\,\cite{leung99,maassen00}).

These solutions are `special' in the sense that they represent \emph{reflectionless} waves and they are the 
only known closed-form exact solutions of the perturbation equations (see Appendix~\ref{sec:special} for details). 
Furthermore, they are associated with the so-called algebraically special QNM 
frequencies\,\cite{chandra84,chandrabook}.

The R-W equation admits a particularly simple special solution (see Appendix~\ref{sec:special} for a derivation):
\be
X_* =  \left ( n + \frac{3 M}{r} \right ) e^{-i\omega_* x},
\label{Xspecial}
\ee
where
\be
\omega_* = -i \frac{n(n+1)}{3M}, 
\label{special}
\ee
are the Schwarzschild algebraically special QNM frequencies. 
It is evident from (\ref{Xspecial}) that this special solution describes a purely ingoing wave
(albeit of imaginary frequency) that suffers no reflection by the black hole wave potential.  

Given that
\be
X_*^{\pp} + [ \,\omega^2_* - V_\rRW (r) \,] X_* = 0,
\ee
we can generate a DT using  (\ref{ueq0}) with 
\be
c = \omega_*^2 - \omega^2, \qquad u = X_*. 
\ee
The resulting `special' $f$-function is,
\be
f_* = i \omega_* + \frac{3M (r-2M)}{r^2 (nr + 3M)},
\label{fstar}
\ee
which is real-valued and regular for any $r >0$. 
 
For the new potential $Q = \omega^2 - V$ we have
\be
V  = V_\rRW + 2 f^\p_*,
\ee
which after some straightforward algebra can be identified as the Zerilli potential $V_\rZ$. 

We thus have the Darboux-generated equation,
\be
Y^{\pp} + [ \,\omega^2 - V_\rZ  \,] Y= 0,
\ee
where
\be
Y = X^\p + f_* X. 
\label{YXeq}
\ee
What remains to be shown is that (\ref{YXeq}) is identical to the Chandrasekhar transformation. 
To that end, we first need to invert (\ref{trans1}). After differentiation and use of the Zerilli equation,
we obtain
\be
Z = \frac{1}{C} \left ( X^\p - \frac{A}{B} X \right ),
\label{trans2}
\ee
where
\be
C = -A (\omega^2 -V_\rZ ) + B^\p - \frac{B^2}{A}
= i  (\, \omega_* - \omega \,), 
\ee
that is, $C$ is in fact a constant parameter\,\footnote{Note that (\ref{trans2}) is regular at $\omega=\omega_*$:
when $C= 0$, we find $ X^\p /X = A/B =  -f_* $ and further that $X = e^{-\int f(x) dx} $. 
This turns out to be exactly the algebraically special solution for the R-W equation [see Eq.~(\ref{u_exact1})].}. 

For the rescaled Zerilli function $\cZ = C Z$ (which of course also solves the Zerilli equation) we have
\be
\cZ = X^\p   -\frac{A}{B}  X.
\label{trans3}
\ee
This should be identical to the Darboux relation (\ref{YXeq}); indeed it is easy to verify that
 \be
f_* = -\frac{A}{B}. 
\ee
This completes the proof of the Chandrasekhar transformation being a special case of the DT
between the R-W and the Zerilli equation (the opposite Zerilli $\to$ R-W transformation 
is generated by $-f_*$, see comment in Section~\ref{sec:darboux}). 

It should be noted that we could have arrived at the same conclusion following a slightly 
different approach, without actually relying on the special solution $X_*$. 
Had we inserted the potentials $q=\omega^2-V_\rRW$ and $Q = \omega^2 - V_\rZ$ in 
the two basic Darboux equations (\ref{feq}) and (\ref{dfeq}) we would have verified that these 
are indeed satisfied. 

This procedure also provides us with the explicit $f=f_*$ function. From that, we can
derive the exact special R-W solution \emph{without} any prior knowledge of it:
\begin{align}
\frac{u^\p}{u} &= -f_* = \frac{A}{B} \quad \Rightarrow \quad u = e^{-\int f_*(x) dx} = X_*,
\label{u_exact1}
\\
\nonumber \\
u^{\pp} &= ( f^2_* - f^{\p}_* ) u  \quad \Rightarrow \quad
u^{\pp} + ( \omega^2_* - V_\rRW ) u = 0.
\label{u_exact2}
\end{align}
According to the analysis of Appendix~\ref{sec:isospectral}, the fact that the R-W and Zerilli potentials
are short-ranged (i.e. they decay faster than $x^{-1}$ as $|x| \to \infty$) and Darboux-related means
that they share identical transmission coefficients and QNM frequency spectra. In fact, as a result 
of the DT's real $f$-function, their reflection coefficients are equal too. Hence, viewed from the Darboux 
theory point of view, there is really nothing `mysterious' about the common properties of Schwarzschild
axial and polar perturbations.

%%%%%%%%%%%%%%%%

\subsection{What is so special about the special solution?}
\label{sec:2special}

If one was tasked with producing a new potential from $V_\rRW$ that is non-singular
and barrier-like what would a suitable choice for the auxiliary function u be?

Given that $u$ solves the R-W equation for some frequency $\omega_0$ it should behave as
a superposition $c_{+}  e^{ i\omega_0 x} + c_{-}  e^{- i\omega_0 x} $ for $|x| \gg M$. 
Then any solution with $\mbox{Re} (\omega_0) \neq 0$ is likely to have
roots at finite $x$-values and, consequently, lead to a singular potential $Q$.  
This problem may persist even if $\mbox{Re}(\omega_0)=0$, unless $c_{+}/c_{-} >0$. 

A secure counter-measure against a singular $Q$ would be a $u$ solution that behaves as a single 
exponential when $|x| \gg M$, in other words, a reflectionless wave. This should be of the form 
$ u = S(r) e^{\pm i\omega_0 x}$ with $S(r)$ a rational function of $r$ with no real-valued roots  
(this form also ensures that the exponential factor does not appear in the $f$ and $Q$ expressions,
thus allowing for $\mbox{Re} (\omega_0) \neq 0$ as indeed happens to be the case in the Kerr
spacetime\, \cite{chandra84})).
 We can see that, in the end, this line of reasoning selects the algebraically special solutions with
$\omega_0 = \omega_*$ as the natural choice for generating a regular DT from the R-W equation.

There are actually two such special solutions. 
As discussed in Appendix~\ref{sec:special} it is possible to derive a second independent 
solution $\tilde{X}_*$ via the method of the Wronskian (as expected, this second solution 
describes a reflectionless outgoing wave of frequency $\omega_*$), see Eq.~(\ref{X2_1}). 
In principle, one can repeat the analysis of this section and use $\tilde{X}_*$ as a DT generator,
hence producing another wave potential for Schwarzschild perturbations. As it turns out, however,
this is an exercise of limited practical value. The second solution  $\tilde{X}_*$ cannot be written in 
closed form with respect to the multipole $n$; even worse, its high degree of algebraic complexity 
leads to a new Darboux potential that is significantly more complicated  than $V_\rRW$ or $V_\rZ$.

As a final remark we show that the DT can be used to derive a special solution $Z_*$ 
of the Zerilli equation from the corresponding R-W solution. Since  for $\omega=\omega_*$
we have  
\be
Z_* = X^\p - \frac{X_*^\p}{X_*} X,
\ee
we can use $X = \tilde{X}_*$ to obtain,
\be
Z_* = \frac{ \tilde{X}^\p_* X_* - X_*^\p \tilde{X}_*  }{X_*}  ~\Rightarrow ~ Z_* = \frac{1}{X_*}.
\ee
where the last step follows from the fact that equations of the form (\ref{yeq0}) have a 
constant Wronskian (which can be omitted since anyway $Z$ is defined up to a multiplicative constant). 
This short calculation is just an example of a more general technique for
finding exact solutions of the Darboux-produced $Y$-equation\,\cite{greekbook}.

%%%%%%%%%%%%%%%%%%%%%%%%%%%%%%%%%%%%%%%%%%%%%%%%%%%%

\subsection{The Bardeen-Press equation and the need for a generalised Darboux transformation}
\label{sec:BP}

Apart from the R-W/Zerilli pair, Schwarzschild perturbations can be fully described by a third
equation, the Bardeen-Press (B-P) equation\,\cite{BP73}. This equation served as the precursor of 
the more general Teukolsky formalism for Kerr perturbations; as a result, the Teukolsky equation 
reduces to the B-P equation in the non-rotating limit.

Its explicit form is,
\be
\frac{d^2 \Phi}{dx^2} + \left (\, \omega^2 -V_{\rm BP} \, \right ) \Phi = 0,
\ee
with 
\be
V_{\rm BP} = -\frac{4 i\omega}{r^2} \left ( r-3M \right ) + \frac{1}{r^3} \left [\, 2 (n+1) (r-2M) + 2M\, \right ].
\label{VBP}
\ee
Unlike $V_\rRW$ and $V_\rZ$ the B-P potential is frequency-dependent,  complex-valued and long-ranged 
(since $V_{\rm BP} \sim 1/r$ at $r \to \infty$). Because of these undesired properties, it is useful to be able to
transform the B-P equation to the R-W/Zerilli equations and vice-versa. Indeed, such transformations 
between $\Phi$ and $X,Z$ were found by Chandrasekhar\,\cite{chandra75}.

As it turns out, the $\Phi \leftrightarrow \{X,Z\}$  relationships  are not of the Darboux type (\ref{darb0}). 
In fact, we can prove quite generally that the DT \emph{cannot} produce short-ranged potentials $Q(x)$ from a
long-ranged potential $q(x)$. 

Assuming rational potentials of the form, 
\begin{align}
	q = \omega^2 - \frac{p_1(x)}{q_1(x)}, && Q = \omega^2 - \frac{p_2(x)}{q_2(x)},
	\label{qQrationals}
\end{align}
where $p_1$, $p_2$, $q_1$ and $q_2$ are polynomials with (by assumption) 
$p_1/q_1 = \mathcal{O}\left(x^{-1}\right) $ and $p_2/q_2 = \mathcal{O}\left(x^{-2}\right)$ as $|x| \rightarrow \infty$. 
We can expand in inverse powers of $x$:
\begin{align}
	q &= \omega^2 - \left(\frac{a_1}{x} + \frac{a_2}{x^2} + \frac{a_3}{x^3} + ...\right),
\label{initpot}	
	\\
	Q &= \omega^2 - \left(\frac{b_2}{x^2} + \frac{b_3}{x^3} + \frac{b_4}{x^4} + ...\right).
	\label{finpot}
\end{align}
Inserting these into the Darboux integral relation (\ref{Qqeq}) 
leads to 
\begin{widetext}
\be
\label{proof}
\frac{a_1}{x^2}+\frac{2(a_2+b_2)}{x^3} + \frac{3(a_3+b_3)}{x^4} + ... 
= \left( \frac{a_1}{x}+\frac{a_2-b_2}{x^2}+\frac{a_3-b_3}{x^3} +... \right)
\left(c + a_1 \ln x - \frac{a_2-b_2}{x} - \frac{a_3-b_3}{2x^2}-...\right).
\ee
\end{widetext}
Because the coefficients of like variables must equal zero to satisfy equation (\ref{proof}), the coefficients 
of the logarithmic term are required to be zero, i.e. $a_1 =0$. Therefore the initial potential, $q$, is required to 
be without the long-range term. 

This proof can be easily adapted to the case of interest here where the potentials $V_{\rm BP}$ and 
$V_\rRW, V_\rZ$ are functions of $r$ instead of $x$. Eqs.~(\ref{qQrationals})-(\ref{finpot}) are written in terms of
$r$ and going from $d/dx$ to $d/dr$ in (\ref{Qqeq}) adds the extra rational factor $dr/dx = 1-2M/r$.
The end result is an equation of the form (\ref{proof}) with $r$ in the place of $x$. 

Having established that the $\Phi \leftrightarrow \{X,Z\}$ transformations do not fall under the ordinary
Darboux theory, it is not surprising that the relevant Chandrasekhar relations are of the form,
\be
Y = \cA(r) \Phi^\p + \cB(r) \Phi, \qquad Y = \{X,Z\}.
\label{chandraBP}
\ee 
where the functions $\cA, \cB$ can be found in Ref.\,\cite{chandra75}.

%%%%%%%%%%%%%%%%%%%%%%%%%%%%%%%%%%%%%%%%%

\subsection{The generalised Darboux transformation}
\label{sec:gdarboux}

The above transformation is a special case of the general scheme, 
\be
Y = \beta(x) y^\p + f(x) y,
\label{GDT}
\ee
a form that appears in the original 1882 Darboux paper\,\cite{darboux}.
It is therefore appropriate to dub it the `generalised Darboux transformation' (GDT).
Obviously, the GDT reduces to the ordinary DT for a constant $\beta$.

Following a similar procedure as in Section~\ref{sec:darboux}, we find that (\ref{GDT}) 
generates a potential $Q$ that is related to the initial potential $q$ via
\be
Q = q - \frac{2f'}{\beta} - \frac{\beta''}{\beta}.
\label{genQ}
\ee
In addition we get a Riccati equation
\be
\beta^2 q + f^2 + \beta' f - \beta f' = c= \mbox{const.}
\ee
Introducing the $u(x)$ function as
\be
f = -\beta \frac{u^\p}{u},
\ee
we get the auxiliary equation
\be
u^{\pp} + \left ( \, q + \frac{c}{\beta^2} \,  \right ) u = 0.
\label{GDTueq}
\ee
Once again, we may find a set of equations for $f$ and $f^\p$: 
\begin{align}
\label{genfp}
	f' &= -\frac{1}{2} \left [ \, \beta(Q - q) + \beta'' \, \right ],
\\
\nonumber \\
\label{genf}
	f &= \frac{\beta(Q + q)' + \beta'(Q + 3q) + \beta'''}{2(Q - q)}.
\end{align}
Requiring compatibility between $f$ and $f'$ gives,
\be
\label{compat}
\frac{\beta(Q + q)' + 2\beta' (Q + q) + \beta'''}{(Q - q)} = - \int dx \beta(Q - q).
\ee
This shows that, as a result of the extra degree of freedom represented by $\beta(x)$,  
there is no `hard' constraint between $q$ and $Q$ (as it was the case for the DT, see Eq.~(\ref{Qqeq})). 
Among other things, it is now possible to transform a long-ranged potential $q$ to a short-ranged 
potential $Q$ by a suitable choice of $\beta$ -- an example is provided by the Chandrasekhar 
transformation (\ref{chandraBP}). As discussed in the following section, because of this extra `handle' 
the GDT has been heavily  employed in the context of Kerr perturbation theory where the starting point
is always the Teukolsky equation with its long-range potential. Crucially, following an analysis similar 
to that of Appendix~\ref{sec:isospectral}, it can be shown that for the general case of rational 
$\beta(r), f(r)$ the GDT preserves the QNM spectrum of the original equation.

%%%%%%%%%%%%%%%%%%%%%%%%%%%%%%%%%%%%%%%%%%%%%%%%%%%

\section{Darboux theory for Kerr black holes}
\label{sec:darbouxK}

\subsection{Perturbation theory in Kerr}
\label{sec:kerr}

Once a black hole acquires spin the distinction between axial and polar parity metric
perturbations ceases to be useful since the two sectors become coupled. In fact, perturbation
theory in the Kerr spacetime is formulated on the basis of Newman-Penrose curvature scalars rather than the metric
itself\,\cite{chandrabook}. The famous Teukolsky formalism encapsulates all the necessary information 
into a single wave-like equation that can be separated with respect to the radial, angular and time 
coordinates\,\cite{teuk73}. 

The resulting radial Teukolsky equation is the formalism's main component -- this is briefly reviewed in 
Appendix~\ref{sec:canonical}. However, this equation has the undesirable property of a long-range potential; 
as a consequence, it does not admit simple plane wave solutions at $|x| \to \infty$ and 
is problematic in numerical implementations.

The problem is remedied by transforming the radial Teukolsky equation to a new wave equation 
with a short range potential. As we have seen in Section~\ref{sec:BP}, this procedure cannot be done 
with the ordinary DT and instead a GDT, $ Y = \beta (x) y^\p + f (x) y$, has to be employed.  

This endeavour was spearheaded in the 1970s by Chandrasekhar \& Detweiler who were able to derive
a tetrad of new Kerr wave equations with short range potentials (which are albeit singular for certain $r$ 
and $\omega$). These equations make direct contact with the R-W and Zerilli equations since they 
reduce to these latter equations in the Schwarzschild limit. 

Nowadays the Chandrasekhar-Detweiler (C-D) equations have been supplanted by another GDT-Teukolsky 
equation derivative, the Sasaki-Nakamura (S-N) equation\,\cite{SN82},\cite{mino97}. The S-N equation features 
a non-singular, short-range potential and reduces to the R-W equation in the Schwarzschild limit. 
Its advantages over the C-D formalism extend to the inhomogeneous version of the equation,
featuring a better-behaved source term.

Given the Darboux connection between the R-W and Zerilli equations it is natural to ask if their Kerr 
generalisations (the C-D equations) are similarly related. This issue is addressed in Section~\ref{sec:CDtetrad}. 
Again motivated by the Schwarzschild analysis, it makes sense to study the role of the Kerr algebraically 
special solutions (these are separately discussed in Appendix~\ref{sec:special}) in the context of the Kerr Darboux theory. 
This is the subject of Section~\ref{sec:Kspecial}.

%%%%%%%%%%%%%%%%%%%%%%%%%%%%%%%%%%%%%%%

\subsection{The Chandrasekhar-Detweiler tetrad of Kerr wave equations}
\label{sec:CDtetrad}

In the 1970s series of black hole perturbation theory papers by Chandrasekhar \& Detweiler\,\cite{cd76, detweiler77}  
the Kerr spacetime dynamics is explored with the help of wave equations that are derivatives of the 
Teukolsky equation. In Ref.~\cite{cd76} this latter equation is first written in a canonical form 
(see Appendix~\ref{sec:canonical}) and is subsequently changed, via a GDT,  to a new C-D equation:
\be
\frac{d\tilde{\cR}^2}{d\tx^2} + \tilde{U}_\rT \tilde{\cR} = 0 ~ \longrightarrow ~
\frac{d^2 Y}{d\tilde{x}^2} + \left ( \omega^2 - V_\rC \right ) Y = 0. 
\label{CD_eq}
\ee 
These equations feature the non-standard tortoise coordinate $\tx$ defined as 
\be
\frac{d\tx}{dr} = \frac{r^2+\alpha^2}{\Delta}, \qquad \alpha^2 = a^2 -\frac{am}{\omega},
\ee
where $a$ is the Kerr spin parameter and $\Delta= r^2-2Mr + a^2$.

By design, the new potential $V_\rC$ has a number of attractive properties not present
in the Teukolsky potential $U_\rT$: it is real (when $\omega$ is real), short-ranged and
depends on $\omega$ only implicitly through the parameter $ \alpha$. 
As a result, $V_\rC$ becomes frequency-independent 
for axisymmetric ($m=0$) perturbations. The explicit form of $V_{\rm C}$ is~\cite{cd76}:
\begin{multline}
V_\rC = \Delta \left [\, \frac{\lambda(\lambda+2)}{q + \beta_2 \Delta} - \beta_2 \frac{\Delta}{\rho^8}
\right.
\\
\left. + \frac{( \kappa_2 \rho^2 \Delta - h )( \kappa_2 \rho^2 q -\beta_2 h)}
{\rho^4 ( q + \beta_2 \Delta)(q - \beta_2 \Delta)^2 } \, \right ],
\label{Vc}
\end{multline}
where
\begin{align}
\rho^2 &= r^2 + \alpha^2,
\\
q &= \lambda \rho^4 + 3\rho^2 (r^2-a^2) - 3 r^2 \Delta,
\\
h &= q^\prime \Delta -\Delta^\prime q,
\end{align}
and
\begin{align}
\beta_2 &= \pm 3 \alpha^2,
\\
\kappa_2 & = \pm \left \{ \,  36 M^2 -2\lambda [ \alpha^2 (5\lambda +6) -12 a^2] \right.
\nonumber \\
 & \left. \quad + 2\beta_2 \lambda (\lambda+2) \, \right \}^{1/2}.
\end{align}
The signs in $\beta_2$ and $\kappa_2$ can be chosen independently which means that  $V_\rC$ 
is in reality a tetrad of Kerr potentials. In the Schwarzschild limit, $a=\alpha= \beta_2=0$ and the 
$\kappa_2 = 6M$ ($\kappa_2 = -6M$) equation reduces to the Zerilli (R-W) equation.

Detweiler~\cite{detweiler77} derived a similar tetrad of potentials but this time for wave equations 
written in terms of the usual tortoise coordinate $x$ (for simplicity we use the same symbol $Y$ 
for the wave function but it should not be confused with~(\ref{CD_eq})):
\be
\frac{d^2 Y}{d x^2} + U_\rD Y = 0, 
\label{D_eq}
\ee
where\,\footnote{Some typographical errors in the original Detweiler potentials 
were subsequently corrected in Ref.~\cite{SI90}.}
\be
U_\rD =  -\frac{K^2}{(r^2+a^2)^2}   + \frac{\rho^4 V_{\rm C} }{(r^2+a^2)^2}  + J^2 
- \frac{\Delta J^\prime}{r^2+a^2} ,
\ee
and
\be
J(r) =  \frac{\Delta}{(r^2+a^2)^2}  \frac{ma r}{\omega \rho^2 }.
\ee
The two potential tetrads coincide for axisymmetric modes, $m=0$, along with $x=\tilde{x}$.  

Given that these four potentials are Kerr generalisations of the Darboux-related R-W and Zerilli potentials, 
it is natural to ask if they are related in a similar way. This can be immediately checked with the help of the 
basic Darboux equations (\ref{feq}) and (\ref{dfeq}) for $f$ and $f^\prime$. 

Remarkably, we find that certain pairs of the original Chandrasekhar-Detweiler potentials $U_\rC = \omega^2 - V_\rC$ 
are indeed Darboux-related. Using the notation $(\kappa_2,\beta_2) $ for labelling different members of the tetrad,
the Darboux-related pairs are:
\be
 (-,\pm) \quad  \longleftrightarrow \quad (+,\mp), 
\label{pairs}
\ee
that is, potentials with opposite signs of $\kappa_2$ and $\beta_2$. Apart from its obvious anti-symmetry,
the relation (\ref{pairs}) reduces to the R-W $\leftrightarrow$ Z relation in the $a=0$ limit. 

Repeating the exercise for the $U_\rD$ potentials reveals that these are \emph{not} Darboux-related 
unless $m=0$ (which is expected since $U_\rD = U_\rC$ in axisymmetry).
This disparity between $U_\rD$ and $U_\rC$ can only be attributed to the use of different tortoise coordinates. 

Resuming the Darboux analysis of the $U_\rC$ potentials, we find that the relevant $f$
function is too messy to be displayed. Instead, we repeat the calculation within a slow-rotation
approximation ($a/M \ll 1$) and derive the much simpler leading-order $f$ function:
\be
f_\rC = f_*  + a m  f_1 + {\cal O}(a^2),
\label{fC}
\ee
where $ f_* $ is the Darboux function between the R-W and Zerilli equations 
(Section~\ref{sec:darbouxRW}, Eqn.~(\ref{fstar})) and 
\be
f_1 = -\frac{(n r+3M)^{-2}}{3 Mr}  \left [\,   \frac{ (n+3)}{(n+1) } \omega \cF_1 
+  \frac{\cF_2}{9 \omega r^3  M^2} \, \right ],
\ee
with
\begin{align}
\cF_1 (r) & =  n^2 (2n+1) r^3 + 6n(2n+1) M r^2 + 18 n M^2 r 
\nonumber \\
&+ 18 M^3,
\\
\nonumber \\
\cF_2 (r)  &=   n^4 (n+1)(5n+3)r^6 +  6n^3 (n+1)(5n+3)M  r^5 
\nonumber \\
&+ 9n^3 (5n+3) M^2 r^4 + 9n(10n^2-18 - 9n) M^3 r^3
\nonumber \\
& + 27 (10 n^2 -9 +15n) M^4 r^2 -162  (n-6) M^5 r 
\nonumber \\
&- 972 M^6.
\end{align}
It can be noticed that $f_\rC$ is regular and contains $\omega$ as a  free parameter -- this implies 
that $f_\rC$, unlike $ f_*$, is not generated by an algebraically special solution. In fact, our discussion
below on the role of the special solutions in the Kerr Darboux theory will reveal that $f_\rC$ could
not have been generated by such a solution.    

As a final remark, we may point out that the Darboux connection between the C-D potentials (\ref{pairs}) entails 
identical QNM spectra. As it turns out, all four potentials are isospectral\,\cite{detweiler77}; this is a direct
consequence of the GDT origin of these potentials from the same Teukolsky equation.

%%%%%%%%
\begin{table*}
\begin{minipage}{135mm}
	\begin{tabular}{rccccc}
	\hline
	Equation  & Potential 1 & Potential 2 & Darboux relation & \quad  Section\\
	\hline\hline
	(\ref{Vz}) $\leftrightarrow$ (\ref{Vrw}) & Zerilli & Regge-Wheeler & DT &  \ref{sec:scptb}\\
    (\ref{Vrw})$\leftrightarrow$ (\ref{VBP}) & Regge-Wheeler & Bardeen-Press & GDT  &  \ref{sec:BP}\\
    (\ref{VBP})$\leftrightarrow$ (\ref{Vz}) & Bardeen-Press & Zerilli &  GDT &   \ref{sec:BP}\\
	(\ref{Ut}) $\leftrightarrow$ (\ref{Usn}) & Teukolsky & Sasaki-Nakamura & GDT &  \ref{sec:BP}\\
	(\ref{Vc}) $\leftrightarrow$ (\ref{Vc}) & Chandra-Detweiler $(-,\pm)$ & Chandra-Detweiler $(+,\mp)$ & DT&   \ref{sec:CDtetrad}\\
%	(\ref{Vc}) $\leftrightarrow$ (\ref{Vc}) & Chandra-Detweiler $(\mp,\mp)$ &  Chandra-Detweiler $(\pm,\pm)$ &  GDT   &\ref{sec:CDtetrad}\\
	\hline 
	\end{tabular}
	\caption{List of black hole potentials and their Darboux relations: DT stands for the ordinary Darboux transformation (Section~\ref{sec:darboux}) 
	and GDT for the generalised Darboux transformation (Section~\ref{sec:gdarboux}). }
	\label{tab:overview}
\end{minipage}
\end{table*}
%%%%%%%%%

%%%%%%%%%%%%%%%%%%%%%%%%%%%%%%

\subsection{Using the Kerr algebraically special solutions}
\label{sec:Kspecial}

The existence of special reflectionless solutions of the perturbation equations is preserved 
when moving from the Schwarzschild to the Kerr spacetime. As discussed in Appendix~\ref{sec:special}, 
starting from a special solution of the Teukolsky equation written in terms of the tortoise coordinate 
$\tx$, we can easily produce special solutions for the other canonical-form Kerr perturbations equations.

These can be collectively denoted as 
\be
y^{\pp}_* + [\, \omega_*^2 - V(\omega_*, r) \, ] y_* = 0,
\label{yspecial}
\ee
where the wave potentials $V$ are all rational functions of $r$ and 
transcendental functions of $\omega$ (as a result of the frequency-dependent angular eigenvalue $\lambda$).   

Comparison of (\ref{yspecial}) against the $u$-equation of the DT, 
\be
u^{\pp} + [ \, q (\omega ,x) +c \,] u = 0,
\ee
reveals that the identification $u = y_*$ is possible for potentials of the form $q = \omega^2 - V (r)$, after
choosing $c = \omega^2_* -\omega^2$. Then $y_*$ can generate a DT -- this was the case for the
R-W equation.  
However, the trick fails for the Kerr potentials $V (\omega, r)$ where $\omega$ enters in a non-additive way
(the same is seen to be true for the B-P potential (\ref{VBP})).  
Therefore, the Kerr special solutions \emph{cannot} generate ordinary DTs. 

The situation changes drastically if we consider the special solutions within the framework of the
GDT. As we have seen, this transformation leads to a $u$-equation of the form
\be
u^{\pp} + \left [ \, q(\omega,x) + \frac{c}{\beta^2(x)} \,  \right ] u = 0.
\ee
The free function $\beta(x)$ can be chosen so that\,\footnote{Note that without loss of generality we can 
set $c=1$ since $\beta$ is defined up to a multiplicative constant.}
\be
\frac{c}{\beta^2(x)} = q (\omega_*,x) - q(\omega,x),  
\label{beta_choice}
\ee
and as a result 
\be
u^{\prime\prime} +q(\omega_* ,x)  u = 0 \quad \Rightarrow \quad u = y_*.
\ee
Therefore, a Kerr special solution can always be used to seed a GDT.

The Kerr special solutions are derived in full detail in Appendix~\ref{sec:special}.
The starting point is the special solution admitted by the $\tx$-Teukolsky equation:
\be
R_* = P(r) e^{i\omega_* \tx},
\label{teukspecial0}
\ee
where $ P= A_3 r^3 + A_2 r^2 + A_1 r + A_0 $.  From this it is straightforward
to derive special solutions for the canonical Teukolsky and S-N equations. 

As a first simple application of the GDT scheme (\ref{beta_choice}) we consider the B-P equation 
(in other words, the $a=0$ limit of the canonical Teukolsky equation (\ref{CD_eq}). 
Without going into much detail we simply report that the GDT leads to a new potential $Q$ that is qualitatively
similar to $q = \omega^2 -V_{\rm BP}$  (i.e. regular, complex-valued, frequency dependent and long-ranged)  
but is a significantly more complicated rational function of $r$. 

As a second application we may consider the canonical S-N equation
\be
\frac{d^2 \cX}{dx^2} + U_{\rm SN} \cX = 0. 
\ee
where the (short-ranged) potential $U_{\rm SN}$ can be found in Appendix~\ref{sec:canonical}.
The associated special solution is
\be
\cX_* =  \frac{2}{r}  \left (\frac{r^2+a^2}{\eta}\right )^{1/2} \left (\, A_1 + \frac{3}{r} A_0 \,\right ) e^{i\omega_* \tilde{x}}.
\ee
The application of the GDT leads to
\begin{align}
\beta &= \left [\, U_{\rm SN} (\omega_*,r) -  U_{\rm SN} (\omega,r) \, \right ]^{-1/2},
\\
f & = - \frac{\beta}{\cX_*} \frac{d\cX_*}{dx}.
\end{align}
The new potential,
\be
Q_{\rm SN} = U_{\rm SN}  -2\frac{f^\p}{\beta} - \frac{\beta^{\pp}}{\beta},
\ee
is short-ranged (since $ (2f^\p  + \beta^{\pp} )/\beta \sim {\cal O}(r^{-3})$ at infinity)
but at the same time is a very complicated function of $r$. 

The above examples illustrate the two main features of the GDT generated by the
Kerr special solutions: a short (long)-range potential $q$ transforms to a short (long)-potential
$Q$ with the final potential being significantly more complicated than the initial one. 
Thus, the conclusion is that the use of algebraically special solutions in the Kerr Darboux theory is of
limited practical value.

%%%%%%%%%%%%%%%%%%%%%%%%%%%%%%%
\section{Concluding discussion}
\label{sec:conclusions}

Our results are summarised in Table~\ref{tab:overview}. It shows which
potentials describing perturbations of a black hole are Darboux related
and which are related only by the generalized Darboux transformation.
Apart from highlighting the role of the DT itself in black hole perturbation 
theory and exploring the role of the algebraically special solutions, we have made a new 
contribution by unveiling the Darboux relation between the C-D Kerr potentials.

It will be noted that certain equations, which are particularly
useful, are related by the GDT. An example is the transformation between the Teukolsky 
and S-N equations which is often used because the short-ranged potential of the latter equation 
is more numerically tractable. We have shown that the GDT is required to relate a long-ranged potential 
to a short-ranged one. In general it would be interesting to study in a systematic way the DT of the 
inhomogeneous black hole wave equations and understand the implications for the source terms.

It is, of course, discouraging that the added complexity of the generalized type of transformations 
must be tackled to realize this useful kind of application. This raises the possibility that useful
transformations such as the aforementioned one between the Teukolsky and S-N equations will in future 
only be found by tour-de-force calculations like the one by Sasaki \& Nakamura, or by some inspired guesswork. 
But it is our hope that connecting this area of research to the broader literature on isospectral potentials 
will facilitate such efforts. It may even be that future efforts should look beyond the Darboux
method. Other methods of seeking isospectral potentials, such as the
one described in Abraham \& Moses\,\cite{abmoses}, exist. It has been shown by Luban \&
Pursey\,\cite{LP86} to be distinct from the classical DT. Another, but similar, approach 
is provided by Pursey in\,\cite{purseyI}. A further paper by the same author discusses using the three techniques 
in combination to find new isopectral equations\,\cite{purseyIII}.
To the best of our knowledge techniques of this sort have
yet to be applied to the study of black hole perturbations.

As a final remark, we speculate on a possible new application of the DT.
It is well known (but not so well understood) that polar and axial QNMs of uniform density relativistic stars 
(the so-called $w$-modes) come near to coincide when the stellar compactness is not too extreme, see Ref.\,\cite{na96}. 
This could be an indication that at some appropriate limit, in addition to the assumed decoupling 
between the spacetime and matter perturbations, the axial and polar wave equations may become
Darboux-related.

We hope to be able to explore some of the issues discussed here in future work.

%%%%%%%%%%%%%%%%%%%%%%%%%%%%%%%%%%%%%%%%%%%

%\acknowledgements

%%%%%%%%%%%%%% APPENDIX %%%%%%%%%%%%%%%%%%%%%%%%%%%%%%

\appendix

\section{Isospectrality of the Darboux transformation}
\label{sec:isospectral}

The DT between a certain class of potentials is characterised by the key property of preserving (with some extra 
assumptions) the transmission and reflection amplitudes between the initial and final potentials $q,Q$. 
In the context of systems like black holes, this property entails isospectral potentials, that is, potentials that 
share a common spectrum of QNM frequencies. 
This is discussed in detail below (a similar analysis can be found in the textbook\,\cite{chandrabook}).  

We consider the Darboux-related equations
\be
y^{\prime\prime} + q(x) y = 0, \qquad  Y^{\prime\prime} + Q(x) Y = 0.
\ee
with the extra assumption of potentials that are non-singular, barrier-like, and of short-range, i.e.
\be
\{ q , Q \} = \omega^2 + {\cal O} \left (\frac{1}{x^2} \right ), \qquad |x| \to \infty,
\ee
where $\omega$ is a constant frequency. As a consequence $y, Y$ will behave as plane-waves 
$ \sim e^{\pm i\omega x} $ at $|x| \to \infty$.

From the Darboux relations $Q = q - 2f^\prime$ and $ q +  f^2 -f^\prime = C^2 =\mbox{const.}$ 
we have,
\be
 f (\pm \infty) = f_0 + {\cal O}  \left (\frac{1}{x} \right ), \qquad \omega^2 + f_0^2 = C^2,
\ee
with $f_0$ constant.

For concreteness, we assume that $y$ describes an incoming wave at $+\infty$ that undergoes
partial reflection and transmission by the $q$ potential. Then,
\bear
&& y(+\infty) = A_{\rm in}  (\omega) e^{-i\omega x} + A_{\rm out} (\omega) e^{i\omega x},
\\
&& y (-\infty) = B_{\rm in} (\omega) e^{-i\omega x}.
\eear
The reflection and transmission coefficients associated with this scattering process are, 
\be
I_{\rm ref} = \frac{ | A_{\rm out} |^2 }{|A_{\rm in} |^2}, \qquad I_{\rm tr} =  \frac{ | B_{\rm in} |^2 }{|A_{\rm in} |^2}.
\ee
The QNM frequencies are defined as the values of $\omega$ that admit a purely outgoing wave at $x=+\infty$ and 
a purely ingoing wave at $x=-\infty$. This requirement is satisfied by the roots of
\be
A_{\rm in} (\omega) = 0 \quad \Leftrightarrow \quad I_{\rm ref}^{-1} = I_{\rm tr}^{-1} =0.
\label{qnms}
\ee
Using the Darboux relation $ Y = y^\prime + f y$ we obtain the asymptotic forms for the
$Y$ function,
\begin{align}
Y(+\infty) &= \left [ \,  -i\omega  +  f(+\infty) \, \right ] A_{\rm in}   e^{-i\omega x} 
\nonumber \\
&+ \left [\,  i\omega +  f(+\infty) \, \right ]  A_{\rm out} e^{i\omega x}  
\nonumber \\
& \equiv  \cA_{\rm in}   e^{-i\omega x} + \cA_{\rm out} e^{i\omega x},
\\
\nonumber \\
Y (-\infty) &= B_{\rm in} \left [ -i\omega +  f(-\infty)  \right ] e^{-i\omega x}  \equiv {\cal B}_{\rm in} e^{-i\omega x}.
\end{align}
The associated reflection/transmission coefficients are:
\begin{align}
{\cal I}_{\rm ref} &= \frac{ | \cA_{\rm out} |^2 }{|\cA_{\rm in} |^2} =  
I_{\rm ref}  \frac{ |  i\omega +  f_0  |^2 }{|  i\omega  -  f_0 |^2}, 
\\
{\cal I}_{\rm tr} &=  \frac{ | \cB_{\rm in} |^2 }{|\cA_{\rm in} |^2} =   
I_{\rm tr} \frac{ | i\omega -  f_0 |^2 }{|  i\omega  -  f_0 |^2} = I_{\rm tr}.
\end{align}
This result demonstrates the equality of transmission coefficients between Darboux-related 
potentials\,\footnote{The equality extends to the reflection coefficients in the case of real parameters 
$q, Q,\omega, C$ which is the common state of affairs in quantum scattering systems.}. 
It also shows that the QNM condition (\ref{qnms}) for the potential $q$ translates to a similar 
condition, $\cA_{\rm in} =0$, for the $Q$ potential. In other words, the QNM spectra of the two potentials 
coincide.

%%%%%%%%%%%%%%%%%%%%%%%%%%%%%%%%%%%%%%

\section{The canonical  Teukolsky and Sasaki-Nakamura equations}
\label{sec:canonical}

In this appendix we construct the canonical forms of the two basic equations
describing gravitational perturbations in the Kerr spacetime, namely, the Teukolsky and
Sasaki-Nakamura (S-N) equations  (for reviews see\,\cite{chandrabook, mino97}).

The standard form of the radial Teukolsky equation is
\be
\frac{d}{dr} \left ( \frac{1}{\Delta} \frac{dR}{dr} \right ) - \frac{V_{\rm T}}{\Delta^2} R = 0,
\label{teuk_original}
\ee
where
\begin{align}
K(r) &= (r^2 + a^2)\omega -am,
\\
\Delta (r) &= r^2 -2Mr + a^2,
\\
V_{\rm T} &= 8i\omega r + \lambda -\frac{K}{\Delta} \left [\, K + 4i(r-M)\,\right ], 
\label{Vt}
\end{align}
and all other parameters having their usual meaning. In terms of the standard 
tortoise coordinate $x$,
\be
\frac{dx}{dr} = \frac{r^2+a^2}{\Delta},
\label{tortoise}
\ee
the Teukolsky equation becomes
\be
\frac{d^2 R}{dx^2} + p \frac{dR}{dx}  - \frac{\Delta V_{\rm T}}{(r^2+a^2)^2}  R =0,
\ee
where
\be
p (r) = \frac{\Delta^3}{(r^2+a^2)^2} \frac{d}{dr} \left ( \frac{r^2+a^2}{\Delta^2} \right ).
\ee
The canonical form of this equation is
\be
\frac{d^2 \cR}{dx^2} + U_{\rm T} \cR = 0,
\ee
where
\begin{align}
R &= \frac{\Delta}{(r^2+a^2)^{1/2}} \, \cR,
\label{cRtoR}
\\
U_{\rm T} = &- \frac{\Delta}{(r^2+a^2)^2} V_{\rm T} 
+ \frac{(2r^2 -a^2) \Delta^2}{(r^2+a^2)^4} 
\nonumber \\
&+ \frac{2\Delta (rM+a^2)}{(r^2+a^2)^3} - \frac{4(r-M)^2}{(r^2+a^2)^2}.
\label{Ut}
\end{align}
A different version of the Teukolsky equation can be produced if instead of $x$ we use 
Chandrasekhar's $\omega$-dependent tortoise coordinate $\tx$:
\be
\frac{d\tx}{dr} = \frac{\rho^2}{\Delta}, \quad \rho^2 = r^2 + \alpha^2, \quad \alpha^2 = a^2 -\frac{am}{\omega}.
\label{tortoiseC}
\ee
The two tortoise coordinates are related as,
\be
\tx = x - \frac{am/\omega}{\rp- r_{-}} \log \left ( \frac{r-\rp}{r-r_{-}}\right ).
\ee
In terms of $\tx$ (\ref{teuk_original}) becomes:
\be
\rho^4 \, \frac{d^2 R}{d\tx^2} + 2 \left  [\,  r\Delta -2(r-M) \rho^2 \, \right ]
\frac{dR}{d\tx} - \Delta V_{\rm T} R = 0. 
\label{teuk_tx}
\ee
It is this particular Teukolsky equation that allows the easiest derivation of the Kerr algebraically
special solution (the calculation is detailed in Appendix~\ref{sec:special}). 

The canonical form of (\ref{teuk_tx}) is
\be
\frac{d\tilde{\cR}^2}{d\tx^2} + \tilde{U}_{\rm T} \tilde{\cR} = 0,
\ee
with 
\begin{align}
 \tilde{\cR} &= \rho ( r^2 + a^2)^{-1/2} \cR,
\\
\nonumber \\
\tilde{U}_{\rm T} &= U_{\rm T} +  \frac{\Delta}{\rho^8}  \frac{a^2-\alpha^2}{(r^2+a^2)^2}  
\left [\, 2r (r^2+a^2) \rho^2 \right.
\nonumber \\
& \left. + \Delta \left \{\, 5r^4 + 2r^2 ( a^2 + \alpha^2) -(a\alpha)^2 \, \right \} \, \right ].
\end{align}
Next, we consider the S-N equation. Its commonly used form is~\cite{mino97} 
\be
\frac{d^2 X}{dx^2} - F \frac{d X}{dx} - U X = 0, 
\ee
where
\begin{align}
 \eta (r) &= \sum_{k=0}^{4} \frac{c_k}{r^k}, \qquad  F (r) = \frac{\eta^\p}{\eta} \frac{\Delta}{r^2+a^2},
 \\
 U &= \frac{\Delta U_1}{(r^2+a^2)^2} + \frac{1}{2} \left (\, \frac{1}{2} p^2 - p F +  \frac{\Delta p^\p}{r^2+a^2} \, \right ).
\end{align}
The original Teukolsky potential $V_\rT$ makes part of the auxilliary function $U_1$:  
\begin{align}
U_1 &= V_\rT + \frac{\Delta^2}{\hat{\beta}} \left [\, \left (\, 2\hat{\alpha} + \frac{\hat{\beta}^\p}{\Delta} \,\right )^\p
-\frac{\eta^\p}{\eta} \left (\, \hat{\alpha} + \frac{\hat{\beta}^\p}{\Delta} \,\right )  \, \right ],
\\
\hat{\alpha} (r) &= -\frac{iK\hat{\beta}}{\Delta^2} + 3 i K^\p + \lambda + \frac{6\Delta}{r^2},
\\
\hat{\beta} (r) & = 2\Delta \left ( -i K + r-M -\frac{2\Delta}{r} \right ).
\end{align}
The complex parameters $c_k$ can be found in Ref.\,\cite{mino97}. 
 
The canonical S-N equation is then found to be,
\be
\frac{d^2 \cX}{dx^2} + U_{\rm SN} \cX = 0. 
\label{sn2}
\ee 
where
\begin{align}
X &= \eta^{1/2}  \cX,
\\
U_{\rm SN} &= -U -\frac{1}{4} F^2 + \frac{1}{2}  \frac{\Delta}{r^2+a^2} F^\prime.
\label{Usn}
\end{align}
 
 %%%%%%%%%%%%%%%%%%%%%%%%%%%%%%%%%%%%%%

 \section{The algebraically special solutions}
 \label{sec:special}
 
One of the remarkable properties of black hole perturbation theory is the existence 
of reflectionless wave solutions, that is, solutions that describe purely outgoing
or purely ingoing waves with respect to the radial direction.
These `special' solutions are the only known closed-form exact solutions of the perturbation equations
and are associated with the so-called algebraically special frequencies $\omega_*$\,\cite{chandrabook}.
In our analysis below of the special solutions  we recover known results but 
in a much simpler way than the original derivation\,\cite{chandra84}, \cite{chandrabook}.  

The simplest derivation of a special solution can be done in the context of the R-W equation
\be
\frac{d^2 X }{dx^2} + \left ( \, \omega^2 - V_\rRW \, \right ) X = 0,
\label{RWeq5} 
\ee
and we sketch it here. We seek an exact solution of the form
\be
X = P(r) e^{i\omega x},
\label{sol1}
\ee
where $P$ is a polynomial. Depending on the sign of $\omega$, this solution can represent purely 
outgoing or ingoing waves.

Upon inserting (\ref{sol1}) in (\ref{RWeq5}) we find,
\be
 r^2 (r-2 M) P^{\prime\prime} + 2r(M+ i r^2 \omega) P^\prime + 2 [ 3M - (n+1) r ] P  = 0. 
 \label{Peq1}
\ee
Trying a power-law solution $P\sim r^s$ at $r \to 0$ and $r\to \infty$ we find $s=\{ -1,3 \}$ and $s = 0$, respectively.
Therefore, a polynomial solution for $P$ should necessarily have the form
\be
P(r) =  C_0 + \frac{C_1}{r},
\ee
with $C_0,C_1$ constants. This solves the above equation provided
\begin{align}
& \frac{C_0}{C_1} = \frac{n}{3M},
\\
& \omega = -\omega_* = i \frac{n(n+1)}{3M},
\label{omspec1}
\end{align}
where in the last equation we show the explicit form of the Schwarzschild algebraically special frequency.

We have thus been able to obtain an exact, purely ingoing, special solution of (\ref{RWeq5}):
\be
X_* = \left ( n + \frac{3M}{r} \right ) e^{-i\omega_* x }.
\label{rw_special}
\ee

A second independent solution can be found via the well known Wronskian method. 
Given that the R-W equation has a constant Wronskian, this second solution is 
(up to a multiplicative constant):
\begin{align}
\tilde{X}_* &= X_* \int \frac{dx}{X_{*}^2} 
\nonumber \\
&= X_{*} \int dr  \frac{r^3}{(nr + 3M)^2}
\left ( \frac{r}{2M} -1 \right )^{4i\omega_* M-1} e^{2i\omega_* r}.
\end{align}
After some experimentation we find that $\tilde{X}_{*}$ 
should have the general form 
\be
\tilde{X}_{*} = \frac{ e^{i\omega_* x}}{r} \left ( \frac{r}{2M} -1 \right )^{-4i\omega_* M} \tilde{P} (r),
\label{X2_1}
\ee
with $\tilde{P}$ a polynomial. As expected, this solution represents a purely outgoing wave.

Demanding that this ansatz solves (\ref{RWeq5})  leads to
the polynomial expression
\be
\tilde{P}_n (r) =\sum_{k=0}^{k_{\rm max}} a_k r^k, \qquad k_{\rm max} =  1 + \frac{4}{3}n(n+1).
\ee
The coefficients $a_k$ satisfy a three-term recurrence relation; this forbids the construction of a 
closed-form expression for an arbitrary $\tilde{P}_n$. We can of course construct  $\tilde{P}_n$ 
for any given value of $n$. The lowest-order solution is  $\ell= n =2 $ and we find,
\begin{align}
\tilde{P}_2 (\bar{r}) &= \bar{r}^9 - \frac{35}{2}\, \bar{r}^8  + \frac{275}{2}\, \bar{r}^7 -640\, \bar{r}^6 
+\frac{31345}{16}\,\bar{r}^5
\nonumber \\
& -\frac{132149}{32}\,\bar{r}^4 + \frac{388255}{64}\,\bar{r}^3 -\frac{388255}{64} \,\bar{r}^2
\nonumber \\
& +\frac{1941275}{512} \,\bar{r} -\frac{1164765}{1024},
\end{align}
where $\bar{r} = r/M$. Higher order polynomials can be readily calculated but the resulting algebraic 
complexity increases very rapidly with $n$. 
 
An exact algebraically special solution of the form (\ref{sol1}) is also known to exist 
in the Kerr spacetime\,\cite{chandra84}. The `privileged' equation admitting such a simple solution 
is the $\tx$-Teukolsky equation~(\ref{teuk_tx}).

Assuming a solution of the form (the $P(r)$ here is not to be confused
with the earlier Schwarzschild polynomial):
\be
R = P(r) e^{i\omega\tx},
\ee
Eq.~(\ref{teuk_tx}) leads to,
\begin{multline}
\Delta P^{\prime\prime} + 2\left [\, i\omega (r^2 + \alpha^2) -(r-M)  \, \right ] \Delta P^\prime
\\
-(\lambda + 6i\omega r) \Delta P = 0.
\label{Peq2}
\end{multline}
A power-law solution $P \sim r^s$ at $r \to 0$ and $ r \to +\infty$
is possible for $s=\{0,1\}$ and $s=3$ respectively. 
 Therefore, the desired polynomial should be of the form
 \be
 P(r) = A_3 r^3 + A_2 r^2 + A_1 r + A_0.
 \ee
 When this expression is inserted back in (\ref{Peq2}) we get a fifth-order polynomial of $r$.  
 Since  all the coefficients of this polynomial must vanish, we get six equations for the five parameters
 $A_0, A_1, A_2, A_3, \omega$ (one of the $A$-coefficients can be chosen arbitrarily since our 
 equation is homogeneous). Although this looks like an underdetermined system, it turns out that this is not 
 the case because three of the equations are not independent. 

The solution of the system first leads to an equation for the frequency:
\begin{multline}
\cS(\omega)=\lambda^2 (\lambda +2)^2 + 144 \omega^2\left [\, M^2 +  (m-a\omega)^2 a^2 \,\right ] 
\\
+ 8\lambda a\omega \left [\, 5\lambda (m-a\omega) + 6(m+a\omega)\, \right ]= 0. 
\label{starob}
\end{multline}
This expression can be identified as the `Starobinsky constant'\,\cite{chandrabook} and its roots 
$\cS(\omega_*)=0$ define the Kerr algebraically special frequencies (note that the $a=0$  limit of this equation 
is solved by the Schwarzschild special frequencies~(\ref{omspec1}); for an arbitrary $a$ the
frequencies need to be computed numerically).

For the $A$-coefficients we find:
\begin{align}
A_3 &= 48 \omega^3_*,  \qquad  A_2 = 24 i \lambda \omega^2_*,
\\
A_1 &= 6 \omega_* \left [\, 12 a\omega_* (a\omega_* -m)  + 12i\omega_* M  \right.
\nonumber \\
& \left. -\lambda(\lambda +2) \, \right ],
\\
A_0 &= i  [\,  12\,i\omega_* M(\lambda +2)   -\lambda (\lambda+2)^2
\nonumber \\
&  -24 a\omega_* (a\omega_* + m) + 28\lambda a\omega_* (a\omega_*-m) \, ].
 \end{align}
We have thus found the purely ingoing Teukolsky algebraically special solution~\cite{chandra84},
\be
R_* = \left (\, A_3 r^3 + A_2 r^2 + A_1 r + A_0 \, \right )  e^{i\omega_* \tx}.
\label{teuk_special} 
 \ee 
Knowledge of an exact $R$-solution translates to having exact solutions for the  
canonical Teukolsky equations of Appendix~\ref{sec:canonical}. 
For instance, for the canonical equation
\be
\frac{d ^2\cR}{dx^2} + U_\rT \cR = 0,
\ee
we find (after using (\ref{cRtoR}))
\be
\cR_* =  \frac{P(r)}{\Delta} (r^2+a^2)^{1/2} e^{i\omega_* \tx}.
\label{cRspecial}
\ee 
In a similar way we can obtain a special solution of the S-N equation. 
The differential relation between the Teukolsky function $R$ and the S-N function $X$ is
(see e.g.~\cite{mino97}):
\be
X = r^2 (r^2+a^2)^{1/2}  \left ( \frac{d}{dr} -\frac{iK}{\Delta} \right )^2 \left \{ \frac{R}{r^2} \right \}.
\label{XR_transf1}
\ee
For  $R =R_*$ this returns
\be
X_* = \frac{2}{r}  (r^2+a^2)^{1/2} \left (\, A_1 + \frac{3}{r} A_0 \,\right ) e^{i\omega_* \tilde{x}}.
\ee
which in turn leads to the following special solution of the canonical S-N equation, 
\be
\cX_* =  \frac{2}{r}  \left (\frac{r^2+a^2}{\eta}\right )^{1/2} \left (\, A_1 + \frac{3}{r} A_0 \,\right ) e^{i\omega_* \tilde{x}}.
\ee
 
%%%%%%%%%%%%%%%%%%%%%%%%%%%%

%%%%%%%%%%%%%%%%%%%%%

\end{document}